\documentclass{article}
\usepackage{spconf,amsmath,graphicx}
\usepackage{amssymb}
\usepackage{xtab}
\usepackage{hyperref}
\usepackage{subfig}

\title{END-TO-END OPTIMIZED SPEECH CODING WITH DEEP NEURAL NETWORKS}
\name{Srihari Kankanahalli}
\address{Zenovia Interactive\\sri@zenovia.io\thanks{This paper is based on independent research that the author conducted for his Master's Thesis at the University of Maryland, under Dr. David Jacobs.}}
\begin{document}
%
\maketitle
\begin{abstract}
Modern compression algorithms are often the result of laborious domain-specific research; industry standards such as MP3, JPEG, and AMR-WB took years to develop and were largely hand-designed. We present a deep neural network model which optimizes all the steps of a wideband speech coding pipeline (compression, quantization, entropy coding, and decompression) end-to-end directly from raw speech data -- no manual feature engineering necessary, and it trains in hours. In testing, our DNN-based coder performs on par with the AMR-WB standard at a variety of bitrates ($\sim$9kbps up to $\sim$24kbps). It also runs in realtime on a 3.8GhZ Intel CPU.
\end{abstract}
\section{Introduction}
\label{sec:intro}

The everyday applications of data compression are ubiquitous: streaming live videos and music in realtime across the planet, storing thousands of images and songs on a single tiny thumb drive, and more. In a way, improved compression was what made these innovations possible in the first place, and designing better and more efficient methods of compression could help expand them even further (to developing nations with slower Internet speeds, for example).

Essentially all modern compression standards are hand-designed, including the most prominent wideband speech coder: AMR-WB \cite{AMR-WB}. It was created by eight speech coding researchers working at the VoiceAge Corporation (in Montreal) and the Nokia Research Center (in Finland) over two years, and it provides speech at a wide variety of bitrates ranging from 7kbps through 24kbps. (For reference, uncompressed wideband speech has a bitrate of 256kbps.)

Recently, deep neural networks have shown an incredible ability to learn directly from data, circumventing traditional feature engineering to produce state-of-the-art results in a variety of areas \cite{nature-review}. Neural networks have seen significant historical interest from compression researchers, but almost always as an intermediate pipeline step, or as a way to optimize the parameters of an intermediate step \cite{nn-image-compression-review}. For example, Krishnamurthy et al. \cite{nn-vq-speech-images} used a neural network to perform vector quantization on speech features; Wu et al. \cite{wu-predictive-nn} used an ANN as part of a predictive speech coder; and Cernak et al. \cite{ann-phonological-vocoding} used a deep neural network as a phonological vocoder.

Our proposal is different in nature from all of these: we reframe the entire compression pipeline, from start to finish, as a neural network optimization problem (along the lines of classical autoencoders). As far as we know, this is only the second published work to learn an audio compression pipeline end-to-end -- the previous being an obscure early attempt by Morishima et al. in 1990 \cite{morishima-ann-speech} -- and the first to compete with a contemporary standard. Cernak et al. \cite{composition-deep-spiking} proposed a nearly end-to-end design for a very-low-bitrate low-quality speech coder in 2016; however, their pipeline still required extraction of acoustic features and pitch (and was also quite complex, composing several different deep and spiking neural networks together). All other related designs we know of employ ANNs as a mere component of a larger hand-designed system.

In the domain of image compression, there has been some interest in training ANN-based systems since the 1990s \cite{jiang-review}, but this has not yielded state-of-the-art results until fairly recently either (starting August 2016, when Toderici et al. trained a neural network model outperforming JPEG \cite{toderici-image-beatjpeg}). Thus, it seems our work is on the cutting edge of both deep learning research and compression research.

\section{Network Architecture and Training Methodology}
\label{sec:architecture}

\begin{figure}[t]
\begin{minipage}[b]{1.0\linewidth}
  \centering
  \centerline{\includegraphics[width=8cm]{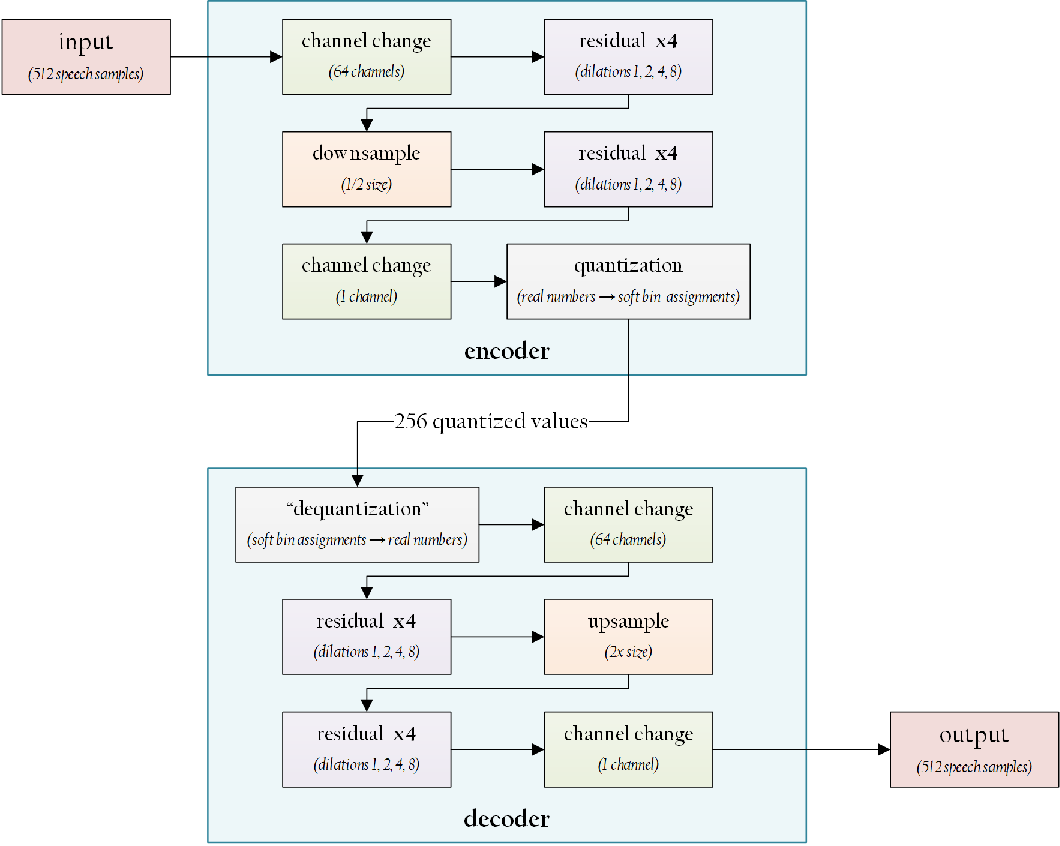}}
\end{minipage}
\caption{Simplified network architecture.}
\label{fig:net_architecture}
\end{figure}

\begin{figure}[tb]

\begin{minipage}[b]{1.0\linewidth}
  \centering
  \centerline{\includegraphics[width=7cm]{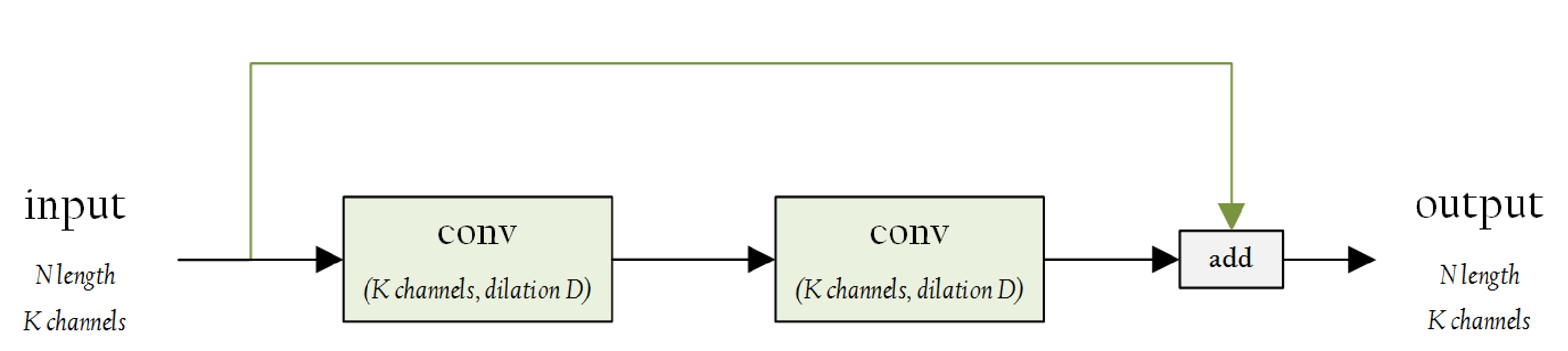}}
  \vspace{0.125cm}
  \centerline{(a) residual \vspace{0.7cm} }
\end{minipage}
\begin{minipage}[b]{1.0\linewidth}
  \centering
  \centerline{\includegraphics[width=7cm]{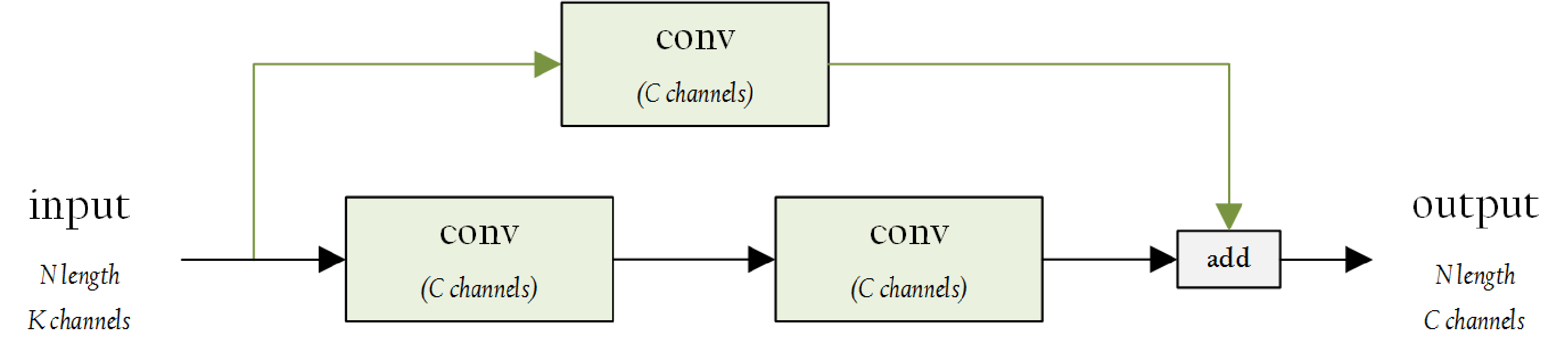}}
  \vspace{0.125cm}
  \centerline{(b) channel change \vspace{0.7cm} }
\end{minipage}
\begin{minipage}[b]{1.0\linewidth}
  \centering
  \centerline{\includegraphics[width=7cm]{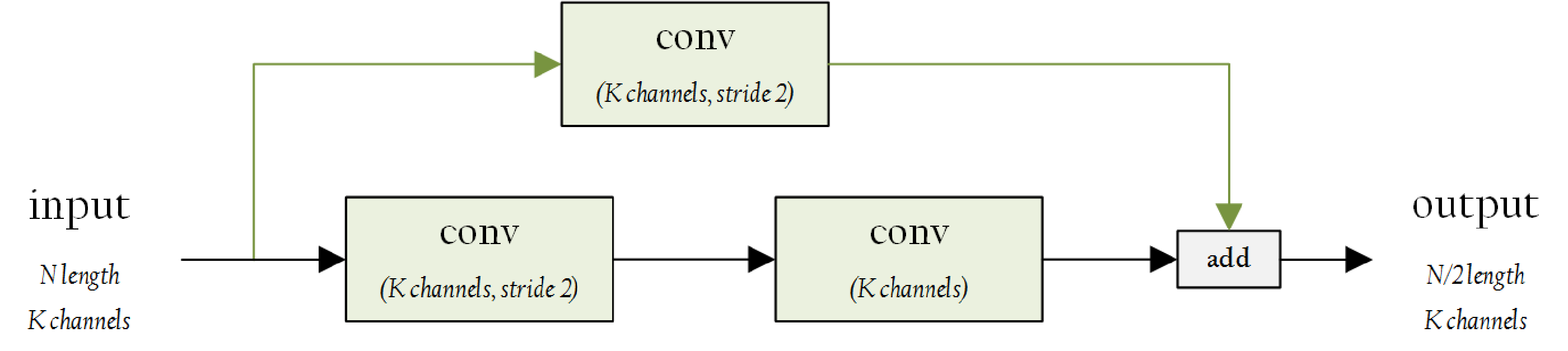}}
  \vspace{0.125cm}
  \centerline{(c) downsample \vspace{0.7cm} }
\end{minipage}
\begin{minipage}[b]{1.0\linewidth}
  \centering
  \centerline{\includegraphics[width=7cm]{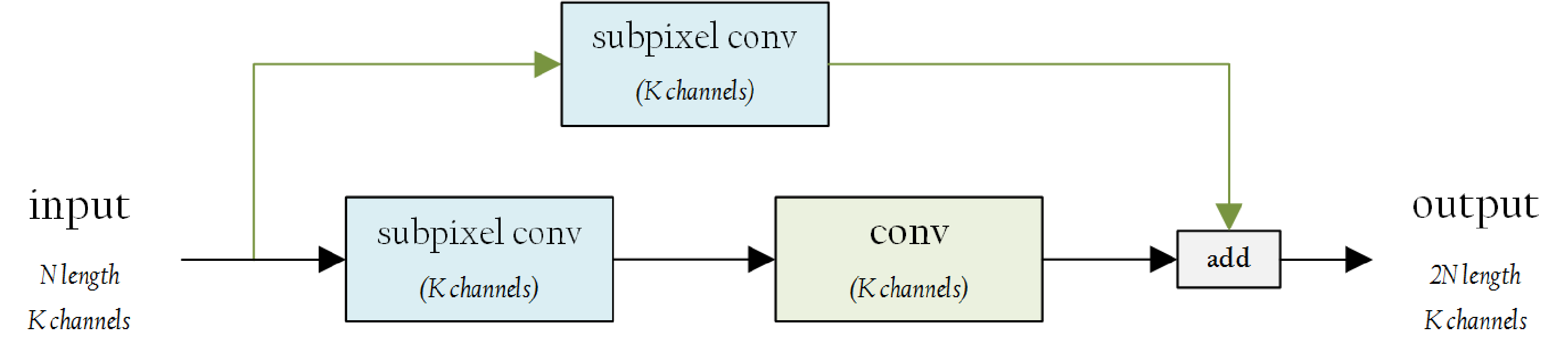}}
  \vspace{0.125cm}
  \centerline{(d) upsample}
\end{minipage}
\caption{The four block types used in our network architecture.}
\label{fig:res_blocks}
\end{figure}

Our network architecture, shown in Figure \ref{fig:net_architecture}, is inspired by both residual neural networks \cite{resnets} and autoencoders. The model is composed of an encoder subnetwork and a decoder subnetwork; it takes in a vector of 512 speech samples (a 32ms speech window) and outputs another vector of 512 speech samples (the reconstructed window after compression and decompression). The network is composed of 4 different types of residual blocks \cite{resnets}, shown in Figure \ref{fig:res_blocks}. All convolutions use 1D filters of size 9 and PReLU activations \cite{prelu}; the upsample block uses subpixel convolutions \cite{subpixel-convolution}. (We were unable to successfully incorporate batch normalization.)

\subsection{Softmax Quantization}

Quantization -- mapping the real-valued output of a neural network into discrete bins -- is an essential part of our pipeline. However, quantization is inherently non-differentiable, and therefore incompatible with the standard gradient-descent-based methods used to train neural networks.

In order to circumvent this, we use a differentiable approximation first discussed by Agustsson et al. \cite{soft-to-hard}. Specifically, we reframe scalar quantization as nearest-neighbor assignment: given a list $B$ of $N$ bins, we quantize a scalar $x$ by assigning it to the nearest quantization bin. This operation still isn't differentiable, but can be approximated as follows:
    \begin{equation}
    D = [ |x - B_1|, ..., |x - B_N| ] \in \mathbb{R}^{N}
    \end{equation}
    \begin{equation}
    S = softmax(- \sigma D)
    \end{equation}
\noindent $S$ is a soft assignment over the $N$ quantization bins, which becomes a hard assignment as $\sigma \to \infty$ (and can later be rounded into one). On the decoder side, we can "dequantize" $S$ back into a real value $\hat{S}$ by taking the dot product of $S$ and $B$. Since Agustsson et al. did not give this approximation a name, we hereby dub it \emph{softmax quantization}.

In practice, we noticed no problems training with very high temperature values from the start. For all experiments, we initialized with $\sigma = 300$, making $\sigma$ and $B$ trainable parameters of the network. (We also found that scalar quantization gave better-sounding results than the vector quantization more prominently discussed by Agustsson et al.)

\subsection{Objective Function}

The network's objective function is as follows:

\begin{equation}
\begin{split}
O(x, y, c) = & \lambda_{mse}\ell_2(x, y) \\ + & \lambda_{perceptual}P(x, y)  \\ + & \lambda_{quantization}Q(c)  \\ + & \lambda_{entropy}E(c)
\end{split}
\end{equation}

\noindent
where $x$ is the original signal,  $y$ is the reconstructed signal, $c$ is the encoder's output (the soft assignments to quantization bins), $\ell_2(x, y)$ is mean-squared error, and $\lambda$ corresponds to weights for each loss. $P(x, y)$, $Q(c)$, and $E(c)$ are supplemental losses, which we now discuss in more depth.

\begin{itemize}
    \item \emph{Perceptual loss.} Training a model solely to minimize mean-squared error often leads to blurry reconstructions lacking in high-frequency content \cite{l2-blurry} \cite{perceptual-metrics}. Therefore, we augment our model with a perceptual loss. We compute MFCCs \cite{mfcc-1} for both the original and reconstructed signals, and use the $\ell_2$ distance between MFCC vectors as a proxy for perceptual distance. To allow for both coarse and fine differentiation, we use 4 MFCC filterbanks of sizes 8, 16, 32, and 128:
    \begin{equation}
    P(x, y)\ =\ \frac{1}{4}\  \sum_{i = 1}^{4}\ \ell_2(M_i(x),\ M_i(y))
    \end{equation}
\noindent where $M_i$ is the MFCC function for filterbank $i$.
    \item \emph{Quantization penalty.} Because softmax quantization is a continuous approximation, it is possible for the network to learn how to generate values outside the intended quantization bins -- and it almost always will, if there is no additional penalty for doing so. Therefore, we define a loss function favoring soft assignments close to one-hot vectors:
    \begin{equation}
    Q(c)\ =\ \frac{1}{256}\ \sum_{i\: =\: 0}^{255} \: [\: (\sum_{j\: =\: 0}^{N - 1}\: \sqrt{c_{i, j}})\ -\ 1.0\: ]
    \end{equation}
    \noindent $Q(c)$ is zero when all 256 encoded symbols are one-hot vectors, and nonzero otherwise.
    \item \emph{Entropy control.} We apply entropy coding to the quantized symbols, which provides a simple way to specify different bitrates without having to engineer entirely different network architectures for each one. Depending on our desired bitrate, we can constrain the entropy of the encoder's output to be higher or lower (by modifying the loss weight $\lambda_{entropy}$ appropriately).

To estimate the encoder's entropy, we compute a probability distribution $h$ specifying how often each quantized symbol appears in the encoder's output, by averaging all of the soft assignments the encoder generates over one minibatch. Thus, our entropy estimate is:
    \begin{equation}
    E(c)\ =\ \sum_{h\: =\: histogram(c)}\ -h_i\: log_2(h_i)
    \end{equation}
\end{itemize}

\subsection{Training Process}

\noindent
We train the network on samples from the TIMIT speech corpuzs \cite{timit}, which contains over 6,000 wideband recordings of 630 American English speakers from 8 major dialects. We create smaller training/validation/test sets from the pre-existing train/test split: our training set consists of 3,000 files from the original train set, our validation set consists of 200 files from the original train set, and our test set consists of 500 files from the original test set. Each set contains an even distribution over the 8 dialects, and they do not share any speakers. Additionally, we preprocess each speech file by maximizing its volume.

We extract raw speech windows of length 32ms (512 speech samples), with an overlap of 2ms (32 samples), using a Hann window in the overlap region. This means that each speech window covers a total of 480 unique samples, or 30ms of speech. Our training process takes place in two stages:

\begin{enumerate}
    \item \emph{Quantization off}. The network is trained without quantization; in this stage, only the $\ell_2$ and perceptual losses are enabled. After 5 epochs, the quantization bins are initialized using K-means clustering, $\lambda_{entropy}$ is set to an initial value $\tau_{initial}$, and quantization is turned on. We found that this "pre-training" period improved the stability and quality of the network's output.
    \item \emph{Quantization on}. The network is trained for 145 more epochs, targeting a specified bitrate. At the end of each epoch, we evaluate the model's mean PESQ over our validation set, and save the best-performing one. We also estimate the average bitrate of the encoder:
    \begin{equation}
    \begin{split}
    bitrate = &\ (windows/sec)\ * \\
              &\ (symbols/window)\ * \\
              &\ (bits/symbol)\ bps \\
    = &\ \frac{16000}{512 - 32}\ *\ 256\ *\ E(c)\ bps 
    \end{split}
    \end{equation}
    If the estimated bitrate is above the target bitrate region, then $\lambda_{entropy}$ is increased by a small value $\tau_{change}$; if it is below the target region, then $\lambda_{entropy}$ is decreased by $\tau_{change}$. This removes the need to manually find the optimal $\lambda_{entropy}$ for each target bitrate. (The target region is defined as our target bitrate $\pm$ 0.45kbps.)
\end{enumerate}

\noindent
During training, we also slowly lower the network's learning rate from an initial value $\alpha_{initial}$ to a final value $\alpha_{final}$, using cosine annealing \cite{sgdr} \cite{shake-shake}. We repeat the training process for each bitrate we want to target; for example, if we want to target 4 different bitrates, we train 4 networks (using the same architecture, but ending up with different sets of weights). The training process takes about 20 hours per network, on a GeForce GTX 1080 Ti.

\section{RESULTS}
\label{sec:results}

\subsection{Objective Quality Evaluation}
\begin{figure}[t]
\centering
\subfloat[training set]{
    \includegraphics[width=3.9cm]{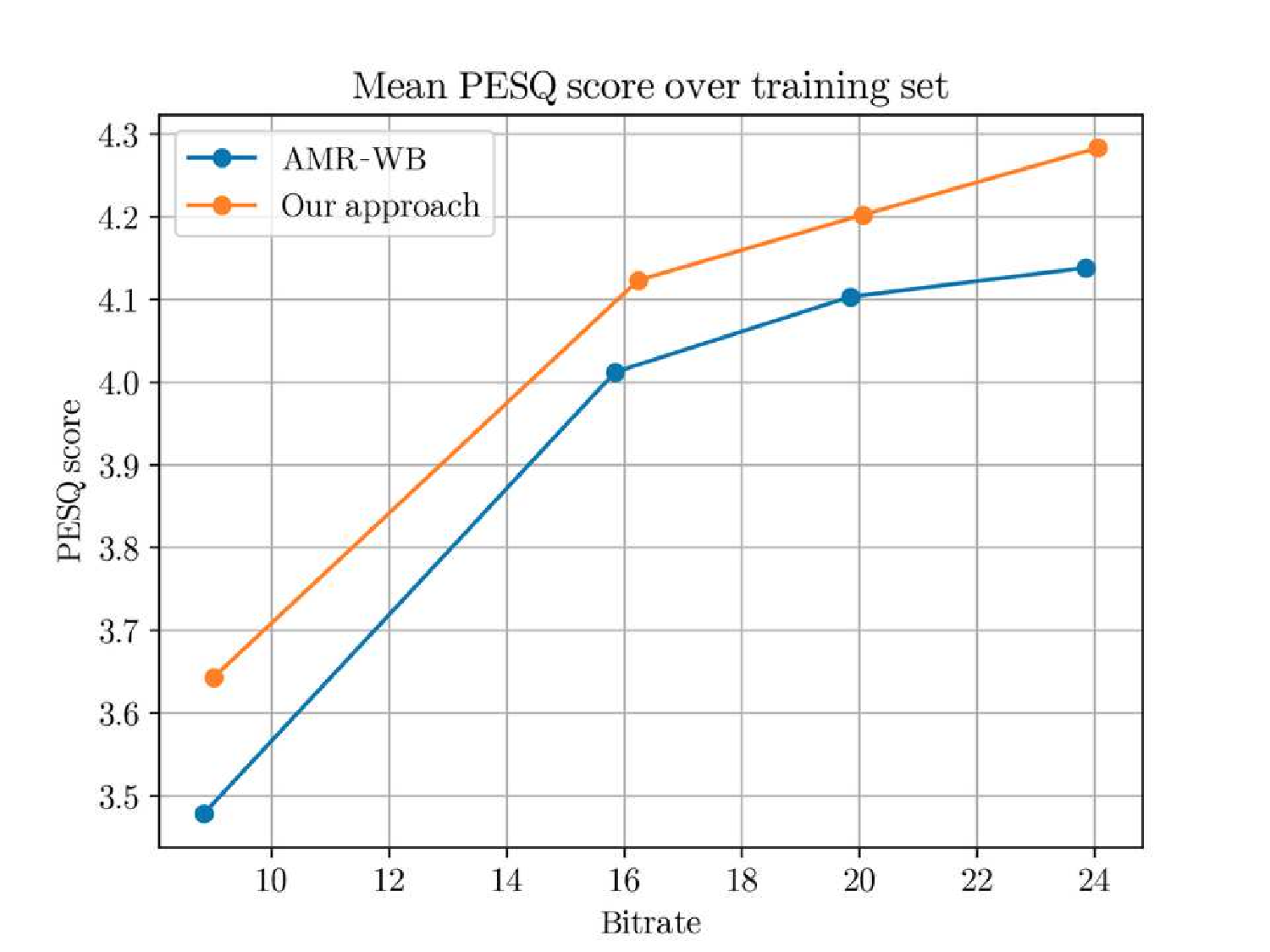}
}
\hspace{0.5em}
\subfloat[validation set]{
    \includegraphics[width=3.9cm]{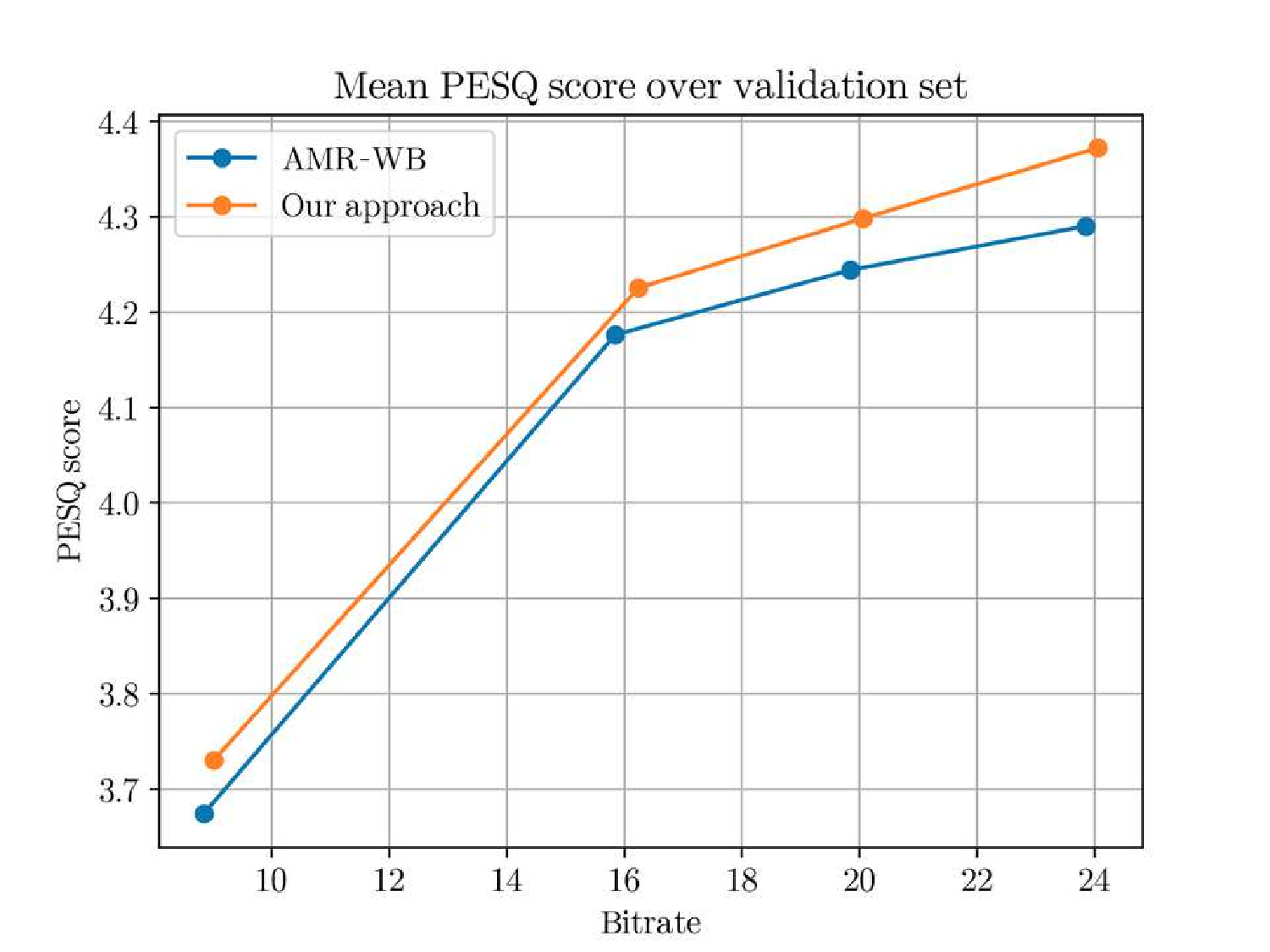}
}

\subfloat[test set]{
    \includegraphics[width=6.0cm]{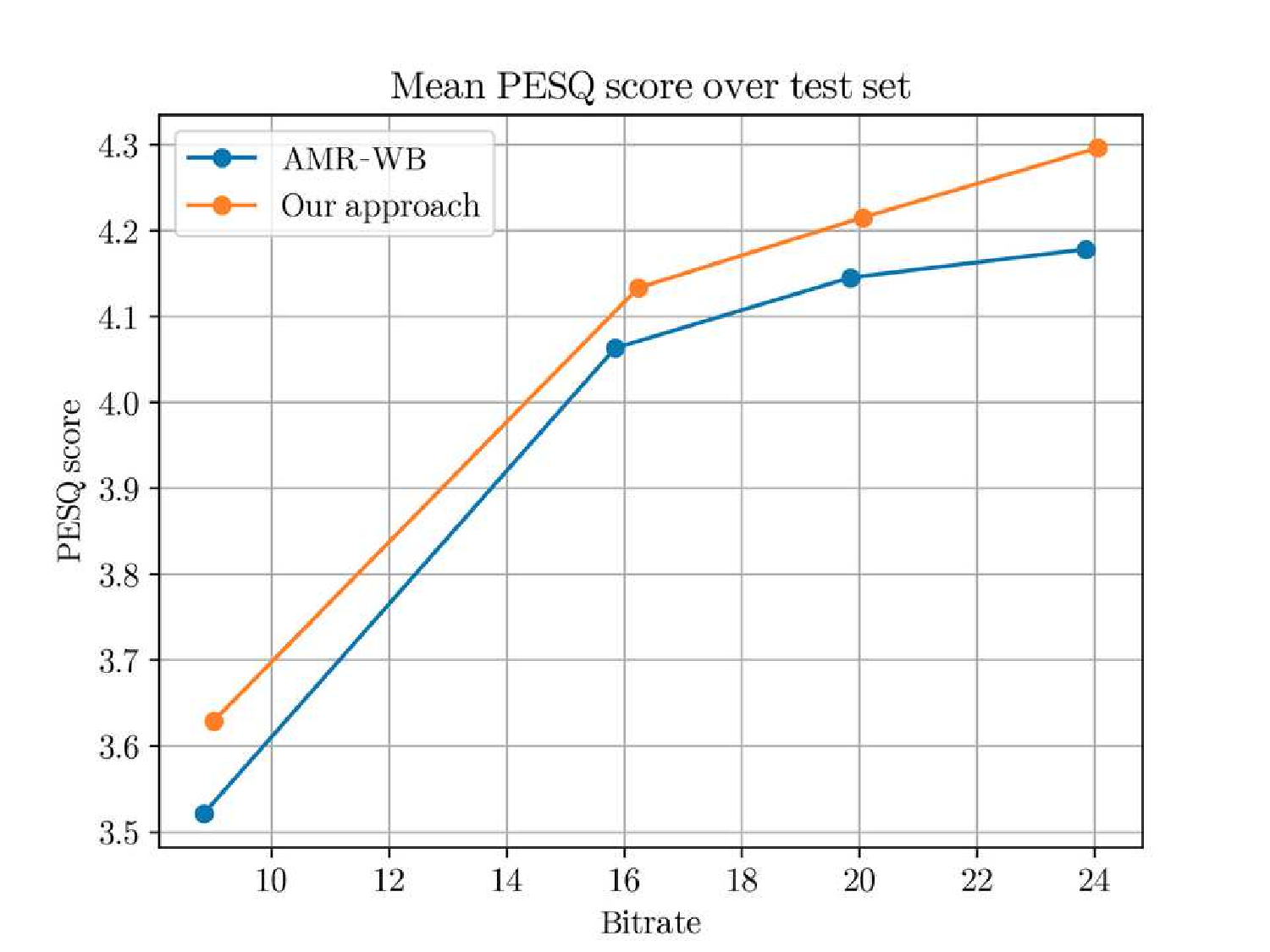}
}

\caption{Mean PESQ of our encoder, compared with AMR-WB at different bitrates.}
\label{fig:pesq_results}
\end{figure}

We evaluated the average PESQ of our speech coder versus the AMR-WB standard around 4 different target bitrates. The results are shown in Figure \ref{fig:pesq_results}, and we reproduce them below:

\begin{center}
    \begin{tabular}{c cc cc}
        \textbf{Dataset} & \multicolumn{2}{c}{\textbf{AMR-WB}} & \multicolumn{2}{c}{\textbf{DNN}} \\
        & \textbf{Bitrate} & \textbf{PESQ} & \textbf{Bitrate} & \textbf{PESQ}\\
        Training set & 8.85 & 3.478 & 9.02 & 3.643\\
        & 15.85 & 4.012 & 16.24 & 4.123\\
        & 19.85 & 4.103 & 20.06 & 4.202\\
        & 23.85 & 4.138 & 24.06 & 4.283\\
        Validation set & 8.85 & 3.674 & 9.02 & 3.730\\
        & 15.85 & 4.176 & 16.24 & 4.225\\
        & 19.85 & 4.244 & 19.70 & 4.298\\
        & 23.85 & 4.290 & 23.71 & 4.372\\
        Test set & 8.85 & 3.521 & 9.02 & 3.629\\
        & 15.85 & 4.063 & 16.24 & 4.133\\
        & 19.85 & 4.145 & 20.06 & 4.215\\
        & 23.85 & 4.178 & 24.06 & 4.296\\
    \end{tabular}
\end{center}

\noindent
Our speech coder outperforms AMR-WB at all bitrates, especially higher rates. The gap is bigger on the training set than on the validation or test sets, indicating possible overfitting (note that we did not use dropout or weight regularization).

\subsection{Subjective Quality Evaluation}

We conducted a simple preference test using Amazon Mechanical Turk. 20 speech files were randomly selected from the test set and processed with both AMR-WB and our method, at the same 4 target bitrates as before. Then, 20 listeners were presented  the original speech signal plus both processed versions (unlabeled and randomly switched). Each listener was asked to pick which of the two he or she preferred. The subjects' average preferences are recorded below:

\begin{center}
    \begin{tabular}{c c c c} 
        \textbf{Target Bitrate} & \textbf{DNN} & \textbf{No Preference} & \textbf{AMR-WB} \\
        9kbps & 25.50\% & 32.00\% & 42.50\% \\
        16kbps & 24.50\% & 37.00\% & 38.50\% \\
        20kbps & 23.50\% & 41.75\% & 34.75\% \\
        24kbps & 23.75\% & 39.00\% & 37.25\% \\
    \end{tabular}
\end{center}

\noindent
Overall, the subjects slightly preferred AMR-WB to our DNN-based coder, with the gap narrowing at higher bitrates. This indicates that more work needs to be done in order to increase our model's subjective quality.

\subsection{Computational Complexity}

We evaluated the average time our model takes to encode and decode one 30ms window, on an Intel i7-4970K CPU (3.8GhZ) and a GeForce GTX 1080 Ti GPU:

\begin{center}
    \begin{tabular}{c c c c} 
        \textbf{Processor} & \textbf{Encoder} & \textbf{Decoder} & \textbf{Total} \\
        CPU & 10.52ms & 10.90ms & 21.42ms \\
        GPU & 2.43ms & 2.35ms & 4.78ms \\
    \end{tabular}
\end{center}

\noindent
Our speech coder runs in realtime (under 30ms for combined encode and decode) without any optimizations beyond those already provided by TensorFlow and Keras. However, it's important to note that real speech coders will need to run on processors much slower than the CPU we used.

\section{CONCLUSION}
\label{sec:conclusion}

We have shown a proof-of-concept applying deep neural networks (DNNs) to speech coding, with very promising results. Our wideband speech coder is learned end-to-end from raw signal, with almost no audio-specific processing aside from a relatively simple perceptual loss; nevertheless, it manages to compete with current standards.

The key to further increasing quality probably lies in our perceptual model, which could be significantly more complex and nuanced. This is where psychoacoustic theory can come into the picture once again: to develop a differentiable perceptual loss for this and other audio processing tasks. In addition, expanding the training data to include music and background noise instead of solely clean speech may be fruitful.

Finally, while our DNN-based coder already runs in realtime on a modern desktop CPU, it's still a far cry from running on embedded systems or cellphones. Model compression, transfer learning, and clever architecture designs are all interesting areas which could be explored here.

\section{HYPERPARAMETERS}
\label{sec:hyperparams}

For purposes of reproducibility, we now make available the list of parameters used for all experiments:

\begin{center}
    \begin{xtabular}{c c} 
        $\sigma_{initial}$ & 300 \\ 
        $\alpha_{initial}$ & 0.025 \\
        $\alpha_{final}$ & 0.01 \\
        $\lambda_{perceptual}$ & 5.0 \\
        $\lambda_{quantization}$ & 10.0 \\
        $\lambda_{mse}$ & 30.0 \\
        $\tau_{initial}$ & 0.5 \\
        $\tau_{change}$ & 0.025 \\
        $N$ & 32 \\ 
        Batch size & 128 \\
        Optimizer & Adam \\
    \end{xtabular}
\end{center}

\noindent
The parameters are listed in roughly descending order by how much manual tuning they required. Source code will be made public after the reviewers' decision. Speech samples are available at: \url{http://srik.tk/speech-coding}

\bibliographystyle{IEEEbib}
\bibliography{sources}

\end{document}